# Efficiency and versatility of distal multisite transcription regulation

Leonor Saiz and Jose M. G. Vilar[*]

Integrative Biological Modeling Laboratory, Computational Biology Program, Memorial Sloan-Kettering Cancer Center, 1275 York Avenue, Box #460, New York, NY 10021, USA

**Corresponding author:**

Jose M.G. Vilar

Computational Biology Program
Memorial Sloan-Kettering Cancer Center
1275 York Avenue, Box #460
New York, NY 10021, USA
Tel: 646-888-2603
Fax: 646-422-0717
E-mail: vilar@cbio.mskcc.org

**Running title:** Multisite transcription regulation

[*] Correspondence: vilar@cbio.mskcc.org

## Summary

**Transcription regulation typically involves the binding of proteins over long distances on multiple DNA sites that are brought close to each other by the formation of DNA loops [1-3]. The inherent complexity of the assembly of regulatory complexes on looped DNA challenges the understanding of even the simplest genetic systems, including the prototypical *lac* operon [4]. Here we implement a scalable quantitative computational approach to analyze systems regulated through multiple DNA sites with looping. Our approach applied to the *lac* operon accurately predicts the transcription rate over five orders of magnitude for wild type and seven mutants accounting for all the combinations of deletions of the three operators. A quantitative analysis of the model reveals that the presence of three operators provides a mechanism to combine robust repression with sensitive induction, two seemingly mutually exclusive properties that are required for optimal functioning of metabolic switches.**

## Results and Discussion

Control of localized events on DNA by proteins bound at distal sites is intimately linked to DNA looping. DNA flexibility plays an important role in mediating long-range interactions [4], allowing proteins bound to non-adjacent DNA sites to come close to each other. This strategy is widely used in eukaryotic enhancers [1] to integrate multiple signals into the control of the transcriptional machinery [1, 2], to the extent that transcription regulation through DNA looping is nowadays considered to be the rule rather than the exception [3, 5]. DNA looping can also be formed by single-protein complexes, including the regulators of many bacterial operons, such as *ara*, *gal*, and *lac* [6], and human proteins involved in cancer, such as retinoic X receptor (RXR) [7] and p53 [8]. The presence of DNA looping with single- and multi-protein complexes is important not only for transcription regulation but also for many other cellular processes, including DNA replication [9], recombination [10], nucleosome positioning [11], and telomere maintenance [12].

To date, there is only limited understanding of the factors that drive macromolecular assembly on looped DNA, especially when multiple binding sites and loops are involved [4, 13]. A notable example is the *lac* operon in *E. coli*, which is still far from being completely understood despite being one of the systems that led to discovery of gene regulation [14].

The *E. coli lac* operon is the genetic system that regulates and produces the enzymes needed to metabolize lactose [14]. The response to lactose is controlled by the *lac* repressor [15] that can bind to $O_1$, the main operator, and prevent the RNA polymerase from binding to the promoter and transcribing the genes. There are also two auxiliary operators, $O_2$ and $O_3$, to which the repressor can also bind but not prevent transcription (Figure 1). Elimination of either one auxiliary operator has only minor effects; yet simultaneous elimination of both of them reduces the repression level by a factor 100 [16]. The reason for this effect is that the *lac* repressor can bind simultaneously to two operators and loop the intervening DNA. Thus, the main operator and at least one auxiliary operator suffice to form DNA loops that substantially increase the ability of the repressor to bind the main operator. Beyond increasing the repression level, it is not clear to what extent DNA looping shapes the properties of transcription regulation and the effects that having three instead of two operators has on the behavior of the system. The lack of a quantitative model of the *lac* operon thus prevents addressing basic general questions about its structure and function.

To incorporate the relevant molecular properties into a quantitative model of transcription regulation, we have used a statistical thermodynamics approach. We consider a decomposition of the free energy of the protein-DNA complex into positional, interaction, and conformational contributions [17]. The positional free energy, $p$, accounts for the cost of bringing the *lac* repressor to its DNA binding site in the protein-DNA complex; interaction free energies, $e$, arise from the physical contact between the repressor DNA binding domains and the different operator sites; and conformational free energies, $c$, account for changes in conformation, including the formation of DNA loops (Figure 1). All these contributions to the free energy can be collected to obtain the free energy $\Delta G(s)$ of a given state $s$ of the protein-DNA complex. The advantage of this approach is that it provides the free energies of a large number of different states from just the individual properties of the interactions and components. Here, different states account for the different ways in which the repressors can bind the three operators.

The free energy of a state is connected to the equilibrium probability $P_s$ of such state through the statistical thermodynamics relationship $P_s = \frac{1}{Z} e^{-\Delta G(s)/RT}$, where $Z = \sum_s e^{-\Delta G(s)/RT}$ is the partition function which serves as normalization factor and $RT$ is the gas constant times the absolute temperature [18].

Straightforward application of the traditional thermodynamic approach [19] in a general framework is of limited use because the number of states that must be considered typically increases exponentially with the number of components [13]. For instance, just the binding of the *lac* repressor to 3 sites would lead to 8 states. If DNA looping is taken into account, the number of states increases to 14. It has become clear recently that it is possible to overcome this limitation and express the free energy of all these states in a compact form by using binary variables [13]. By extending this approach to consider multiple loops in the *lac* operon, we obtain

$$\Delta G(s) = (p+e_1)s_1 + (p+e_2)s_2 + (p+e_3)s_3$$
$$+ (c_{L12} - ps_1s_2)s_{L12} + (c_{L13} - ps_1s_3)s_{L13} + (c_{L23} - ps_2s_3)s_{L23}$$
$$+ \infty(s_{L12}s_{L13} + s_{L12}s_{L23} + s_{L13}s_{L23}),$$

where $s_1$, $s_2$, and $s_3$ are binary variables that indicate whether ($s_i = 1$, for $i = 1, 2, 3$) or not ($s_i = 0$, for $i = 1, 2, 3$) the repressor is bound to O$_1$, O$_2$, and O$_3$, respectively; and $s_{L12}$, $s_{L13}$, and $s_{L23}$ are variables that indicate whether ($s_{Lij} = 1$, for $ij = 12, 13, 23$) or not ($s_{Lij} = 0$, for $ij = 12, 13, 23$) DNA forms the loops O$_1$-O$_2$, O$_1$-O$_3$, and O$_2$-O$_3$, respectively. The subscripts of the different contributions to the free energy have the same meaning as those of the corresponding binary variables. In this case, with 3 interaction and 3 conformational free energies, it is possible to obtain the free energy of 14 states for different repressor concentrations. The dependence on the repressor concentration, $n$, enters the free energy through the positional free energy: $p = p_0 - RT \ln n$, where $p_0$ is the positional free energy at 1 M concentration. An important advantage of the binary variable description is that it can straightforwardly implement "logical conditions". For instance, the infinity in the last term of the free energy implements that two loops that share one operator cannot be present simultaneously by assigning an infinite free energy to those states.

The expression of the *lac* operon is strongly determined by the occupancy of the different operators. Explicitly, transcription is completely abolished when the repressor is bound to $O_1$; otherwise, transcription takes place either at an activated maximum rate $\tau_{max}$ when $O_3$ is free or at basal reduced rate $\chi \tau_{max}$ when $O_3$ is occupied. This reduction by a factor $\chi$ arises because binding of the repressor to $O_3$ prevents the Catabolite Activator Protein (CAP) from activating transcription [20]. Activation is achieved when CAP bound to cyclic AMP binds between $O_3$ and $O_1$ and stabilizes the binding of the RNA polymerase to the promoter [21, 22]. In terms of binary variables, the transcription rate $\tau(s)$ can be expressed as

$$\tau(s) = \tau_{max}(1-s_1)(\chi s_3 + (1-s_3)),$$

which provides a mathematical expression for the observed cis-regulatory transcription control [20, 23].

The concise mathematical expressions $\Delta G(s)$ and $\tau(s)$ completely specify the thermodynamic and transcriptional properties of the system. In particular, the repression level, defined as the maximum transcription over the actual average transcription rate, is given by $R = \tau_{max}/\bar{\tau}$, where the average transcription rate follows from $\bar{\tau} = \frac{1}{Z}\sum_{s}\tau(s)e^{-\Delta G(s)/RT}$. Thus, with this approach, it is possible to obtain a compact description for both the DNA-repressor complex and the control of transcription that scales linearly with the number of regulatory elements [13].

Our model accurately reproduces the observed behavior of the *lac* operon system in quantitative detail over five orders of magnitude of the repression level for three repressor concentrations and eight strains with all the possible combinations of operator deletions (Figures 2 and 3) [20]. The values of the parameters used are the same as those previously reported [4] except for two small variations. One of the changes is a 0.5 kcal/mol shift in the affinity of the repressor for the operators. This difference falls within the typical variation for different experimental conditions. The other change is a decrease of 0.9 kcal/mol in the free energy of forming the $O_1$-$O_3$ loop. This decrease is consistent with stabilization of the loop by binding of CAP to its DNA site between $O_1$ and $O_3$ [24], which was not present in the experiments used to infer the free energies of looping [25, 26]. We have also observed that the experimental repression levels are better accounted for if the deletion of $O_1$ is considered not to be complete. Explicitly, we infer that the binding of the repressor is not completely abolished but reduced by ~5 kcal/mol, which is consistent with the mutation of just three base pairs of the operator [20]. This strong reduction of affinity in a single operator setup would be indistinguishable from a complete deletion, but it is not so in a multi-loop configuration.

In the *lac* operon, long-range interactions are not strictly required and an increase in repression could have been achieved with a single stronger operator or a stronger repressor, which both exist for the *lac* operon [14, 27]. To what extent do the properties of transcription regulation depend on the molecular complexity that multiple DNA loops bring about?

The analysis of the model reveals that the three-operator setup provides an efficient mechanism to combine robust repression with sensitive induction (Figure 4). A key element is CAP, which controls the stability of one of the DNA loops besides activating transcription. When CAP is bound to its DNA site, as in the experiments considered here, the multi-loop system is sensitive to changes in the repressor concentration. The resulting sensitivity would make the system ready for induction when the repressor is inactivated by inducers such as allolactose or IPTG [14]. This result strongly contrasts with previous studies of regulation by single loops that show that repression is highly insensitive to changes in repressor concentration [28]. Such robustness is recovered when CAP is not bound to DNA and the $O_1$-$O_3$ loop is not stabilized. Therefore, repression relies only in the $O_1$-$O_2$ loop and the system displays a lack of sensitivity to changes in repressor concentration. A three-operator system is thus able to put two apparently contradictory properties such as robustness and sensitivity together into a functional metabolic switch.

The statistical thermodynamics approach we have implemented naturally incorporates the underlying molecular complexity into gene regulation models and provides an avenue to accurately infer the effects of multiple DNA loops between different DNA sites. In the *lac* operon, escalating complexity from one to two operators introduces stronger repression; and from two to three operators, concurrent robustness and sensitivity. Thus, the presence of multiple repeated distal DNA binding sites, far from being just a remnant of evolution or a backup system as often assumed [14], can confer subtle, yet important, properties that are not present in simpler setups. These results indicate that key design principles that have been shown to play important roles in shaping the structure of biochemical networks [29-31] are also operating at the molecular level.

**Figure Legends**

**Figure 1: Operators controlling expression of the lacZ, lacY, and lacA genes in the *lac* operon.**

(**A**) Location of the main ($O_1$) and the two auxiliary ($O_2$ and $O_3$) operators, shown as orange rectangles on the thick black segment representing DNA. Binding of the *lac* repressor to $O_1$ prevents transcription of the three *lacZYA* genes.

(**B**, **C**) The bidentate repressor can bind to any of the three operators and simultaneously to any two of them by looping the intervening DNA, resulting in different protein-DNA complexes. Two of the three possible loops are shown: (**B**) one *lac* repressor (shown in red) loops DNA by binding simultaneously to $O_1$ and $O_3$ (loop *L13*) whereas another repressor binds to $O_2$; (**C**) only one repressor is bound to DNA, forming a loop between $O_1$ and $O_2$ (loop *L12*). In both cases, the *lacZYA* genes are repressed. The different contributions to the free energy of the *lac* repressor-DNA complexes, which include positional ($p$), interaction ($e_1$, $e_2$, and $e_3$), and conformational ($c_{L12}$, $c_{L13}$, and $c_{L23}$) contributions, are explicitly indicated in these two cases.

**Figure 2: Model *versus* experimental repression levels.**
The repression levels obtained from the model (see main text) are plotted against their corresponding experimental values [20], showing an excellent quantitative agreement over 5 orders of magnitude for WT and seven mutants accounting for all the combinations of deletions of the three operators. Three different repressor concentrations are considered, which lead to a total of 24 data points, shown as blue squares with different shades indicating increasing repressor concentrations from WT (darkest) with 10 repressors per cell, to 50 (medium shade) and 900 (lightest) repressors per cell. The red line corresponds to the identity between model and experimental values. The values of the parameters used are: $e_1$=-27.6 kcal/mol, $e_2$=-26.105 kcal/mol, $e_3$=-23.925 kcal/mol, $c_{L12}$=23.35 kcal/mol, $c_{L13}$=22.05 kcal/mol, $c_{L23}$=23.65 kcal/mol, $p_0$=15 kcal/mol, and $\chi$=0.03. A deleted operator is modeled by increasing its free energy by 5 kcal/mol.

**Figure 3: Repression level as a function of the repressor concentration for WT and seven mutants accounting for all the combinations of deletions of the three *lac* operon operators.**

For each of the eight cases, the results of the model (red curves) as a function of the concentration of repressors are compared with the experimental data (shaded blue squares) available for three concentrations corresponding to 10 (WT cells), 50, and 900 repressors per cell. The particular set of WT or deleted (X) operators is indicated in each curve with $O_3$-$O_1$-$O_2$ corresponding to the WT *lac* operon and X-X-X, to the mutant with all three operators deleted. The excellent agreement indicates that the model not only captures the repression values quantitatively but also the shapes of the curves, which are very different depending on the mutant. The values of the parameters are the same as in figure 3.

**Figure 4: Robust repression and sensitive induction.**

The repression level with (red curve) and without (blue curve) CAP bound to its DNA site is plotted as a function of the repressor concentration. Without CAP, in addition to a reduced transcription $\tau = \frac{1}{Z}\sum_{s} \tau_{max} \chi (1-s_1) e^{-\Delta G(s)/RT}$, the formation of the $O_1$-$O_3$ loop is ~1 kcal/mol more costly ($c'_{L13}$=23.05 kcal/mol) than with CAP, which leads to an almost flat profile of the repression level around wild type repressor concentrations (gray circle). With CAP, repression is reduced at the same time that the system becomes sensitive to changes in repressor concentration. Variability in repressor concentration over a population of cells, with a distribution such as that shown in green, would lead to a much wider distribution of repression levels with CAP (in red) than without CAP (in blue).

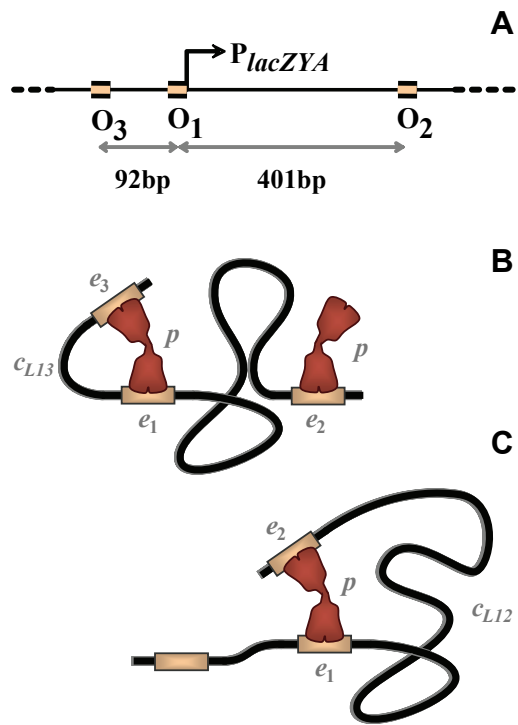

**Figure 1**

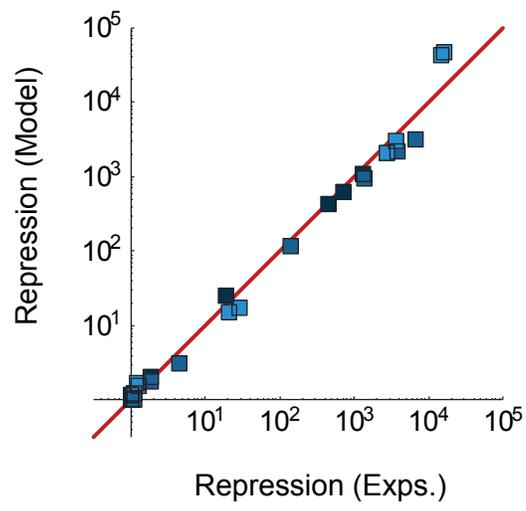

**Figure 2**

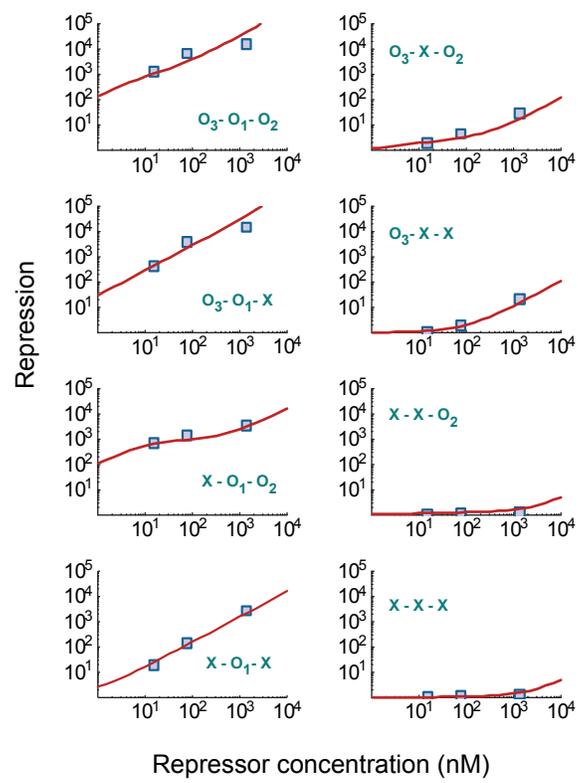

**Figure 3**

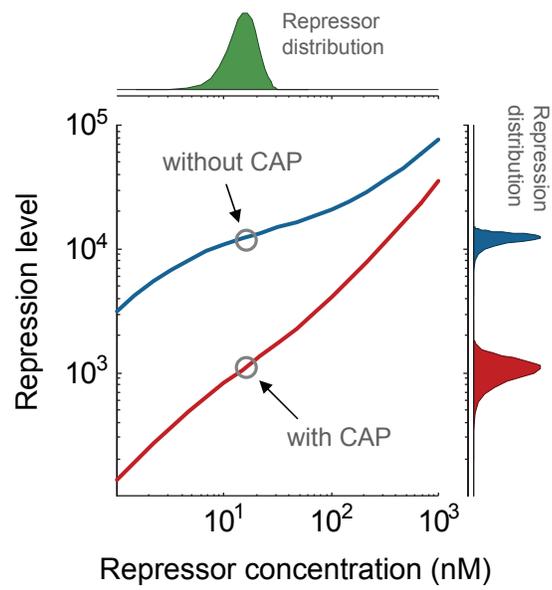

**Figure 4**